% This is the FINAL version
% Thur Jun 22

\input uiucmac.tex
\PHYSREV
\def\Re{\hbox{Re}}
\def\Rel{\hbox{Re$_{\lambda}$}}
\def\dll{D_{LL}(r)}
\def\ebarr{(\bar\epsilon r)^{2/3}}
\def\ck{C_K}
\def\lnRe{\ln\,\Re}
%\unnumberedchapters
%\tolerance 2000
\nopubblock
\titlepage
%
% Comment out the \singlespace below for journal version - default
% is preprint version.
%
\singlespace

\title{\bf Does Fully-Developed Turbulence Exist? Reynolds Number Independence
versus Asymptotic Covariance
}

\author{G.I. Barenblatt$^{1,2}$\foot{Permanent address: Dept. of Applied
Mathematics and Theoretical Physics, University of Cambridge, Silver
St., Cambridge CB3 9EW, U.K.} and Nigel Goldenfeld$^{2}$}
\address{$^1$Department of Theoretical and Applied Mechanics,
University of Illinois at Urbana-Champaign, 104 S. Wright St.,
Urbana, Il. 61801, U.S.A.}
\address{$^2$Department of Physics and Beckman Institute,
University of Illinois at Urbana-Champaign, 1110 West Green Street,
Urbana, Il. 61801-3080, U.S.A}
%\andaddress{}

%\andauthor{ }
%\address
% pagebreak for journal copy
\vfil
%\vfil\eject
\abstract\noindent
By analogy with recent arguments concerning the mean velocity profile of
wall-bounded turbulent shear flows, we suggest that there may exist
corrections to the 2/3 law of Kolmogorov, which are proportional to
$(\ln\,\Re)^{-1}$ at large Re.  Such corrections to K41 are the only
ones permitted if one insists that the functional form of statistical
averages at large Re be invariant under a natural redefinition of Re.
The family of curves of the observed longitudinal structure function
$D_{LL}(r, \Re)$ for different values of Re is bounded by an envelope.
In one generic scenario, close to the envelope, $D_{LL}(r, \Re)$ is of
the form assumed by Kolmogorov, with corrections of $O((\lnRe)^{-2})$.
In an alternative generic scenario, both the Kolmogorov constant $C_K$
and corrections to Kolmogorov's linear relation for the third order
structure function $D_{LLL} (r)$ are proportional to $(\ln\,\Re)^{-1}$.
Recent experimental data of Praskovsky and Oncley appear to show a
definite dependence of $C_K$ on Re, which if confirmed, would be
consistent with the arguments given here.

\bigskip\noindent
Pacs Numbers: 47.27.Gs

\endpage

{\bf\chapter{Introduction}}
\noindent
%\section{}

The term {\it fully-developed turbulence\/} traditionally refers to a
unique state of turbulent behaviour believed to occur for sufficiently
large but finite Reynolds number (Re).  This state is characterised by
local isotropy and homogeneity and associated universal behaviour of
statistical properties, such as moments of the longitudinal velocity
difference $v_r\equiv (\VEC v(\VEC x + \VEC r) - \VEC v(\VEC x))\cdot
\VEC r/|\VEC r|$.  The first theoretical description along these lines
was given by Kolmogorov and Obukhov in 1941 (referred to as
K41)\rlap.\Ref\kolmo{A.N. Kolmogorov \journal Dokl.  Akad. Nauk.  SSSR
&30&301(41); \journal ibid.  &31&99(41)\ [English translation in
\journal Proc. Roy. Soc. London.  Ser. A &434&9(91)]; A.M. Obukhov,
\journal Dokl. Akad. Nauk SSSR &32&22(41).}\ The assumption that such a
limiting state exists, and may be found at large but finite Re, is
non-trivial, and in our view, has not properly been established
experimentally.  The purpose of this note is to investigate how a
breakdown of this assumption would be manifested.

The mathematical expression of the assumption of a limiting state of
fully-developed turbulence is that statistical averages of the flow
exhibit complete similarity\Ref\baren{G.I. Barenblatt, {\sl Similarity,
Self-similarity and Intermediate Asymptotics\/} (Consultants Bureau, New
York, 1979).} with respect to Re.  To explain this statement, let us
consider the second order longitudinal structure function $\dll \equiv
\left<(v_r)^2\right>$.  In K41, $\dll$ has the form
$$
\dll = C_K (\bar\epsilon r)^{2/3},\eqn\kfourone
$$
where $r$ lies in the inertial range, and $\bar\epsilon$ is the mean
rate of energy dissipation per unit mass.  Kolmogorov's form for $\dll$
is based upon both dimensional considerations, and assumptions about
limiting behaviour.  Dimensional analysis shows that the form of $\dll$
must be given by
$$
\dll = \ebarr F(\Re, r/L), \eqn\dllform
$$
where $F(x,y)$ is a universal function to be determined, $L$ is the
external or integral scale and $r$ is always considered to lie in the
inertial range.  Kolmogorov assumed that in the limits $x\rightarrow
\infty$ and $y\rightarrow 0$, the function $F(x,y)$  simply takes the
constant value $\ck$.  In other words, there is {\it complete
similarity\/} with respect to the variables Re and $r/L$.

The existence of the limit of $F(x,y)$ as $y\rightarrow 0$ has been
questioned\REFS\intermm{L.D. Landau (unpublished); see, for example,
L.D. Landau and E.M. Lifshitz, {\sl Fluid Mechanics}, 2nd ed.
(Pergamon, New York, 1987), p. 140.}\REFSCON\obhuk{A.M. Obukhov,
\journal J. Fluid Mech. &13&77(62).}\REFSCON\kosixii{A.N. Kolmogorov,
\journal J. Fluid Mech. &13&82(62).}\refsend\ due to intermittency ---
fluctuations of the energy dissipation rate about its mean value
$\bar\epsilon$.  {\it Incomplete similarity\/} in the variable $r/L$
would require the non-existence of a finite and non-zero
limit of $F(x,y)$ as
$y\rightarrow 0$, and leads in the simplest case to the form
$$
\dll = C_K (\bar\epsilon r)^{2/3} \left(\frac{r}{L}\right)^{\alpha}, 
\eqn\intermittt
$$
where $\alpha$ is the so-called intermittency exponent, believed to be
small and non-negative.

In the present note, we argue that there may be an alternative way,
which we term {\it asymptotic covariance}, in which a lack of complete
similarity can occur.  We try to use physical arguments to constrain the
mathematical form that this might take.  Asymptotic covariance in Re
would imply that there is no unique limiting state of fully-developed
turbulence.  Instead, the manner in which statistical averages evolve
with Re for large Re is governed by a functional form that in the
simplest case is universal.

Precisely the same set of arguments can be made for a seemingly
different, but related problem: the mean velocity profile in a
wall-bounded turbulent shear flow. There is a well-known (and probably
superficial) analogy\Ref\tenn{See \eg, H. Tennekes and J.L. Lumley, {\sl
A First Course in Turbulence\/} (MIT, Cambridge, 1990), pages 147 and
263.}  between the boundary layers and universal scaling regimes of both
the spatial structure of wall-bounded turbulent shear flows and the
local structure of turbulence.  For example, consider turbulent flow in
a pipe.  The viscous wall layer is analogous to the dissipative range in
fully-developed turbulence; the velocity profile (conventionally
described by the universal von K\'arm\'an-Prandtl logarithmic law)
outside the viscous wall layer, but on scales much smaller than the pipe
radius is analogous to the inertial range (conventionally described by
K41); and the non-universal finite-size effects on the flow associated
with scales of order the pipe radius are analogous to the non-universal
behaviour of fully-developed turbulence at the integral scale.  This
problem was already considered in detail\rlap,\Ref\barenjfm{G.I.
Barenblatt, \journal C.R. Ac.  Sci. Paris &313&307(91); G.I.
Barenblatt, \journal J. Fluid Mech.  &248&513(93); G.I. Barenblatt and
V.M. Prostokishin, \journal J.  Fluid Mech. &248&521(93).} and it was
shown that the existing data do not exclude the possibility there of
asymptotic covariance in Re.

This paper is organised as follows.  In section 2, we review the
analysis of the wall-bounded shear flow.  In particular, we propose a
principle of {\it asymptotic covariance}, in which we insist that the
functional form of statistical averages at large Re be invariant under
redefinition of Re.  This implies that a form of incomplete
similarity must occur in terms of the variable $\lnRe$.  In section 3,
we consider the analogous arguments for the local structure of
fully-developed turbulence.  We conclude in section 4 with a brief
discussion of experimental data and some final comments.

{\bf\chapter{Wall-bounded turbulent shear flows}}

We consider a wall-bounded shear flow which is statistically steady and
homogeneous in the longitudinal direction.  Its properties vary only in
the lateral direction, perpendicular to the wall.  A classic example,
which we shall always have in mind, is the flow in a pipe far from the
entrance and outlet.

\section{Mean velocity profile}

Von K\'arm\'an\Ref\karman{Th. von K\'arm\'an, {\sl Mechanische
\"Ahnlichkeit und Turbulenz}, Nachrichten Ges. Wiss. Goettingen,
Math.-Phys.Kl. (1930), pp.58-76.} and Prandtl\Ref\prandtl{L. Prandtl,
\journal Ergebn. Aerodyn. Versuchsanstalt., G\"ottingen, &4&18(32).}
obtained the law for the variation of the mean longitudinal velocity
$\bar u$ in an intermediate region of the turbulent shear flow, outside
a small ``viscous" sub-layer near the wall.  Within this sub-layer, the
stress due to molecular momentum transfer is comparable in magnitude
with that due to turbulent momentum transfer by vortices.  The
universal, Reynolds number independent von K\'arm\'an-Prandtl
logarithmic law has the form
$$
\phi=\frac{1}{\kappa}\ln\,\eta + \tilde C,\eqn\vkp
$$
where
$$
\phi\equiv \bar u/u_*,\quad \eta\equiv u_*y/\nu,\quad
u_*\equiv(\tau/\rho)^{1/2}, \eqn\defsss
$$
$y$ is the distance from the wall, and $\rho$ and $\nu$ are the fluid
density and kinematic viscosity respectively.  The constants $\kappa$
(the von K\'arm\'an constant) and $\tilde C$ are universal according to
the logic of the derivation.  The logarithmic law follows from a strong
assumption of complete similarity, namely that in the intermediate
region the contribution of the molecular viscosity and the external
length scale (\eg, the diameter of the pipe) could be completely
neglected.

It was shown in ref. \barenjfm\ that this assumption is questionable,
and an alternative relationship was proposed, corresponding to
incomplete similarity:
$$
\phi=C\eta^{\alpha},\eqn\alteqnn
$$
where the coefficients $C$ and $\alpha$ may be expressed in terms of the
small parameter
$$
\epsilon\equiv \frac{1}{\lnRe}\eqn\epsdeff
$$
by the expansions
$$
\alpha=\frac{3}{2}\epsilon + O(\epsilon^2), \quad C= 
\frac{1}{\sqrt{3}\epsilon} + \frac{5}{2} + O(\epsilon).\eqn\expanss
$$
It is important that the power $\alpha$ in \alteqnn\ depends upon
the Reynolds number for two reasons.  First, it is well-known that a
Re-independent power law form is inconsistent with the
data\rlap.\Ref\hinze{See, for example, the discussion by J.O. Hinze,
{\sl Turbulence} 2nd. edition (McGraw-Hill, New York, 1987),
pp.629-632.}\

Second, Eyink\Ref\eyink{G.  Eyink, private communication.}  has argued
that a Re-independent power law is inconsistent with the rigorous bound
on the energy dissipation rate given by Doering and
Constantin\rlap,\Ref\doering{G.R.  Doering and P.  Constantin, \journal
Phys. Rev. Lett. &69&1648(92).} (making the plausible assumption that
their bound can be carried over to the pipe geometry), but the form
given in \expanss\ is consistent with the bound.  Eyink's argument is as
follows: the average dissipation per unit mass $\bar\epsilon$ is
$$\bar\epsilon = \frac{1}{L}\int_0^L \bar\epsilon'(y)\,dy, 
\quad \bar\epsilon'(y)={u_*}^2\frac{\pd \bar u}{\pd y},\eqn\defavdid
$$
which for the von K\'arm\'an-Prandtl law \vkp\ can be estimated to be
$$\bar\epsilon=\frac{U^3}{L}\frac{1}{\kappa (\lnRe)^2}, \quad \Re\rightarrow\infty.\eqn\convcase$$
For the velocity profile \alteqnn, however, it can be shown that
$$
\bar\epsilon'(y) \sim \left(\frac{L}{y}\right)^{1-\alpha}
\left[2^\alpha \alpha(\alpha+1)(\alpha+2)\right]^{(3+\alpha)/(1+\alpha)}
 \Re^{-2\alpha/(1+\alpha)}.
\eqn\eyinkss
$$
If $\alpha$ were Re-independent, the dissipation in the intermediate
region would vanish sharply as $\Re\rightarrow\infty$.  However,
a more physically plausible alternative is that there is only a
weak dependence of the dissipation on Re for $\Re\rightarrow \infty$:
$$
\Re^\alpha \sim \hbox{constant} \quad\hbox{\ie, } \alpha \sim \frac{\hbox{constant}}{\lnRe}.\eqn\niceresult
$$
The corresponding result for the dissipation is that
$$\bar\epsilon\sim\frac{U^3}{L}\frac{1}{ (\lnRe)^2}, \quad \Re\rightarrow\infty,\eqn\bcase
$$
which is very similar to the estimate for the von K\'arm\'an-Prandtl
law.

It seems that the scaling law \alteqnn\ manifests a lack of
universality, by virtue of its dependence on Re.  However, this is not
so, in the following sense.  Instead of the traditional universal
straight line in the $\phi$--$\ln\,\eta$ plane corresponding to \vkp,
there is a one-parameter family of curves \alteqnn\ occupying a certain
region of the plane.  This region is nevertheless universal, in the
sense that it is bounded by the envelope of the family, which is a
universal curve.  The equation for the envelope is obtained by
eliminating $\lnRe$ from \alteqnn, written in the form
$\phi=F(\ln\,\eta, \lnRe)$, and the tangency condition $\pd
F/\pd\lnRe=0$, and is found to have the universal form
$$
\phi = 5\left(\sigma^{-1} + \frac{1}{2}\right)\eta^{\sigma\sqrt{3}/10},\quad \sigma\equiv \left(1+\frac{20}{\sqrt3\ln\,\eta}\right)^{1/2}-1.\eqn\univvf
$$
This envelope is in fact close to \vkp\ even for moderate values of
$\ln\,\eta$ and Re, if $\kappa\sim 0.4$ and $\tilde C=5.1$.  For presently
unachievably large values of Re and $\ln\,\eta$, the universal envelope
\univvf\ assumes the form
$$\phi=\frac{\sqrt{3}e}{2}\ln\,\eta + \frac{5e}{2},\eqn\unachi
$$
where the coefficient $2/\sqrt{3}e\approx 0.424$ is close to the
generally accepted value of the von K\'arm\'an constant $\kappa$, but
$5e/2\approx 6.79$ is larger than the generally accepted value of
$\tilde C\approx 5.1\hbox{---}5.5$.

Thus, even when there is a lack of similarity, it seems that for
$\lnRe\gg 1$ in the intermediate region of the shear flow, the state is
described by the relation $\phi=F(\ln\,\eta, \lnRe)$, universal in the
sense that the same function $F$ applies to all turbulent shear flows.

\section{Asymptotic covariance}

It is instructive that the universal relation \alteqnn\ -- \expanss\
contains Re only through the dependence on $\lnRe$.  In fact, this is
inevitable, and a consequence of what we propose to call {\it asymptotic
covariance}.  Asymptotic covariance provides a general constraint on the
way in which lack of complete similarity in Re may be exhibited in a
turbulent flow.  Let us assume that we consider a simple turbulent flow,
such as that in a pipe, and that a putative state of isotropic,
homogeneous turbulence is present on some scale $r\ll L$, where $L$ is
taken to be the diameter of the pipe.  The {\it nature\/} of the
turbulent state should be insensitive to small changes in the
cross-sectional average input and output flow rate $U$ or the diameter
$L$.  Thus, we require that at sufficiently large Re, the {\it
functional form\/} of statistical averages, such as $\dll$ or $\phi$,
should not be influenced by a redefinition of Re such that
$$\Re\rightarrow \Re '\equiv
Z\Re = \Re + \delta \Re, \qquad \delta\Re/\Re\rightarrow 0.\eqn\redeff
$$  
For example,
at sufficiently large Re, we could use for $L$ the radius of the pipe
instead of the diameter, or for $U$ the maximum velocity instead of the
average one \etc, without changing the functional form of $\dll$ or
$\phi$.  Let us suppose that we consider such a statistical average,
whose dependence on Re (and $\eta$ in the case of the shear flow
considered in this section) is through its dependence on a function
$F(\ln\, \eta, \psi(\Re))$ having for definiteness a uniformly bounded
first derivative, and with $\psi$ an unboundedly growing function of its
argument, to be determined by the following considerations.

For any $Z>0$
$$\eqalign{
\psi(\Re)=\psi(\Re_0)+\int_{\Re_0}^{\Re}
\frac{d\psi(\Re')}{d\Re'}\,d\Re'\cr
\psi(Z\Re)=\psi(Z\Re_0)+\int_{\Re_0}^{\Re}
\frac{d\psi(Z\Re')}{d\Re'}\,d\Re'.\cr}
\eqn\defascovv
$$
Here, $\Re_0$ is some reference value of Re.  The first term on the
right hand sides of \defascovv\ at large Re is small in comparison with
the second term, because $\psi$ is unbounded.  However,
$d\psi(Z\Re)/d\Re=Z\psi'(Z\Re)$ where $'$ denotes differentiation with
respect to the argument.  Asymptotic covariance is equivalent to the
statement that
$$
\psi'(\Re) = \psi'(Z\Re).\eqn\meannns
$$
Thus, the right hand side of \meannns\ is simply
$Z\psi'(Z\Re)$, and
$$
\Re\,\psi'(\Re)=Z\, \Re \psi'(Z\Re)\eqn\zmean
$$
giving $\psi(\Re)\propto\lnRe$.  

Let us now consider the question of whether one could use the Taylor
microscale Reynolds number $\Rel$, which is often assumed to vary
approximately as $\sqrt{\Re}$.  From the point of view of advanced
similarity methods and dimensional analysis, the important point about
Re is that it represents a characterization of the system which can be
made {\it a priori}.  That is, it is not an emergent property of the
flow (\ie, depending upon the {\it solution\/} of the equation of
motion), but a property of the {\it constraints\/} or {\it boundary
conditions\/} placed upon the flow.  On the other hand, $\Rel$
represents the response of the flow, and in general, for arbitrary Re,
will not necessarily have a unique, universal dependence on
Re\rlap.\Ref\condmat{ An analogous situation arises in condensed matter
systems subject to an external magnetic field $\bf H$; in such cases, it
is usually not appropriate to work with the magnetic induction $\bf B$,
which includes also magnetization $\bf M$, \ie, the response of the
system to $\bf H$.}\  However, for an asymptotically covariant theory,
at very large Re, the variation of $\ln \,\Rel$ with $\lnRe$ may be
written in the form
$$
\frac{d\ln \,\Rel}{d\lnRe} = a_0+\frac{a_1}{\lnRe} +O(1/(\lnRe)^2)
\eqn\relams
$$
where $a_0$, $a_1$ etc. are constants, so that
$$
\Rel\sim\Re^{a_0}(\lnRe)^{a_1}\eqn\relamform
$$
showing that there is a unique power law relationship between $\Rel$ and
Re.  

{\bf \chapter{Local structure of fully-developed turbulence}}

As explained in the Introduction, Kolmogorov's form \kfourone\ for 
$$\dll = \ebarr F(\Re, r/L)\eqn\dlformagain
$$
is based upon both dimensional considerations, and assumptions about
limiting behaviour.  Let us now examine the question of the lack of
complete similarity of \dllform, based upon considerations analogous to
those used above for wall-bounded turbulent shear flows.

We will assume that the spatial fluctuations in the energy flux diminish
at higher Re, because the flow configurations or processes which
correspond to these fluctuations become increasingly dense throughout
the flow.  In this picture, the scaling assumed in K41 becomes more and
more accurate at very high Re.  We will refer to this assumption as {\it
asymptotic Kolmogorov scaling}\rlap.\Ref\lvov{Note that asymptotic
Kolmogorov scaling occurs in a recent diagrammatic analysis by 
V.S. L'vov and V.V. Lebedev, \journal JETP Lett. &59&577(94);V.S.
L'vov and I. Procaccia, \journal Phys. Rev. Lett. &74&2690(95).}

The behaviour of $F(x,y)$ as $x\rightarrow \infty$ is
the central issue on which we focus.  The manner in which $F$ could fail
to attain a finite limit is constrained by asymptotic covariance,
which implies that
$$
\dll= \ebarr G(\lnRe, r/L),\eqn\newform
$$
for some function $G$ to be determined, and intermittency, \ie\ 
incomplete similarity of $G(x,y)$ in the limit $y\rightarrow 0$.
Conventionally, it is assumed that the manner of violation of K41 is
that $G(x,y) \sim y^{\alpha}$ as $y\rightarrow 0$, where $\alpha$ is an
intermittency correction, assumed to be Re-independent and estimated
experimentally to be positive.  In contrast, we wish to investigate the
consequences of assuming that as $ x\rightarrow\infty$ and $y\rightarrow
0$
$$
\dll = \ebarr A(\lnRe) \left(\frac{r}{L}\right)^{\alpha (\lnRe)} 
\eqn\assump
$$
where the prefactor $A$ and intermittency exponent $\alpha$ depend on
$\lnRe$.  In particular, the assumption of asymptotic covariance
implies that we can write
$$
\alpha = \alpha_0 + \frac{\alpha_1}{\lnRe} + O((\lnRe)^{-2});\eqn\allp
$$
further assuming asymptotic Kolmogorov scaling implies that
$\alpha_0=0$.  Such a form is consistent with the experimental
results of Castaing \etal\Ref\castaing{B. Castaing,
Y. Gagne and E.J. Hopfinger, \journal Physica D &46&177(90).}\  As
before, instead of a universal straight line in the $\dll/\ebarr$ -
$(r/L)$ plane, we obtain a one parameter family of curves, occupying a
certain portion of the plane.  The boundary of this family is a universal,
Re-independent curve, which satisfies both \assump\ and the the
condition $\pd\dll/\pd\Re=0$, \ie
$$
\frac{dA}{d(\lnRe)} = A(\lnRe) \frac{\alpha_1 \ln\,(r/L)}{(\lnRe)^2}.
\eqn\aeqnn
$$
Empirically it is found that the intermittency correction $\alpha$ is
positive, so that $\alpha_1 > 0$.  However, in the inertial range, $r\ll
L$, so that $dA/d(\lnRe) < 0$.  Thus, we may expand $A$ in the small
parameter $\epsilon\equiv1/\lnRe$
$$
A(\lnRe) = A_0 + {A_1}{\epsilon} + O(\epsilon^{-2}), \eqn\aexp
$$
where $A_0$, $A_1$ are non-negative constants.  There are two cases to
consider: (a) $A_0\neq 0$, and (b) $A_0 = 0$.  We will briefly discuss
the experimental estimates\Ref\sreeni{For a survey of available data,
see A.S. Monin and A.M. Yaglom {\sl Statistical Fluid Mechanics}, Vol. 2
(MIT, Boston, 1975); a more recent discussion is given by K.R.
Sreenivasan, {\sl On the universality of the Kolmogorov constant},
preprint.} of $\ck$ in the following section, but for now, we remark
that both the scatter and the recent report of a systematic dependence
on Re\Ref\prask{A. Praskovsky and S. Oncley, \journal Phys.  Fluids
&6&2886(94).} encourage us to consider not only the conventional case
(a) but also case (b).

In case (a), both intermittency corrections to the exponent $\alpha$ and
to the Kolmogorov constant $A_0$ vanish logarithmically as
$\lnRe\rightarrow\infty$.  More precisely, the relation \aeqnn\ gives
$$
\frac{\alpha_1 \ln\,(r/L)}{\lnRe} = -\frac{A_1}{\lnRe}
\frac{1}{[A_0 +  \frac{A_1}{\lnRe} + O((\lnRe)^{-2})]} \eqn\soli
$$
so that the envelope is given by
$$
D_{LL}^{\hbox{env}}(r) = \ebarr\, A_0\, (1 + O(\ln\,(r/L)).\eqn\denvv
$$
Note the correction term in this formula: the cancellations that occur
to first order in $\epsilon$ make the deviations from K41 very
difficult to detect.  Indeed, near the envelope, individual plots of
$\dll$ at fixed but large Re will exhibit corrections to K41 only of
order $O(\epsilon^2)$.

In case (b), the results are more dramatic.  The envelope condition
\aeqnn\ becomes
$$
\frac{dA}{d(\lnRe)} = -\frac{A}{\lnRe}\eqn\abecomes
$$
so that we obtain the condition:
$$
1 = -\frac{\alpha_1\ln\,(r/L)}{\lnRe}.\eqn\condbfix
$$
The universal boundary in the $\dll/\ebarr$ - $(r/L)$ plane is then
represented by the curve
$$
\frac{\dll}{\ebarr} = -\frac{A_1}{e\alpha_1 \ln (r/L)} .\eqn\univboundd
$$
The equation of a member of the family of curves at fixed Re, near the
envelope is
$$
D_{LL}(r) = \ebarr\, \frac{A_1}{e\,\lnRe}, \eqn\envbb
$$
which is K41, but with a Kolmogorov constant $C_K = A_1/e\,\lnRe_\lambda$. 

The phenomenological considerations given above also have implications
for higher moments.  For example, consider the third moment
$D_{LLL}(r)\equiv \left<(v_r)^3\right>$, which is related to $\dll$ by
Kolmogorov's relation
$$
D_{LLL}(r) = -\frac{4}{5}\bar\epsilon r + 6\nu\frac{d\dll}{dr}.
\eqn\vkhk
$$
Using the form \assump\ for $\dll$, and assuming case (b) above, we find
that for large $\lnRe$
$$
D_{LLL}(r) = -\frac{4}{5}\bar\epsilon r \left(1 - \frac{5A_1}{\lnRe}
\left(\frac{\eta}{r}\right)^{4/3} e^{-3\alpha_1/4} +
O((\lnRe)^{-2})\right).  \eqn\dllres
$$
The correction to Kolmogorov's ``4/5 law" is Reynolds number dependent,
in contrast to the situation in K41.

{\bf \chapter{Discussion}}

Sreenivasan\refmark{\sreeni} has recently surveyed the available data on
$C_K$, for different flows and Re, and came to the conclusion that there
is no Reynolds number dependence for the data taken as a whole.
However, there is considerable scatter in the data, and as noted by
Sreenivasan, a controlled series of measurements in a {\it single\/}
flow geometry for a wide range of values of Re has yet to be performed.

To our knowledge, the only measurements which explore the systematics of
the possible Reynolds number dependence of $C_K$ are those of Praskovsky
and Oncley\rlap,\refmark{\prask}in which a weak dependence of $C_K$ on
Re is reported.  These authors attempted to fit this dependence with a
power law, but, as shown in \FIG\figi{The Kolmogorov constant, as
measured by Praskovsky and Oncley\rlap,\refmark{\prask} plotted on a
logarithmic scale against Taylor microscale Reynolds number.  Also shown
are best fits to the functional form given by \aexp, for the two cases
$A_0\neq0$ and $A_0=0$.  Inset (A) shows the scatter when
$C_K\lnRe_\lambda$ is plotted against Re$_\lambda$.  Inset (B) shows the
scatter when $\Delta C_K\lnRe_\lambda$ is plotted against Re$_\lambda$,
where $\Delta C_K\equiv C_K - 0.45$.}figure \figi, their data are
consistent with the logarithmic forms proposed here.   Figure \figi\
shows the data, together with two best fits to the form given by \aexp,
for the two cases $A_0\neq0$ and $A_0=0$.  The former case, with a
non-zero limit of $C_K$ as $\Re\rightarrow\infty$, may be a slightly
better fit than the latter.  One measure of this is shown in the insets
of figure \figi: Inset (A) shows the scatter when $C_K\lnRe_\lambda$ is
plotted against Re$_\lambda$, whilst inset (B) shows the scatter when
$\Delta C_K\lnRe_\lambda$ is plotted against Re$_\lambda$, where $\Delta
C_K\equiv C_K - 0.45$.  A slight upwards trend with increasing
Re$_\lambda$ may be discernable in (A), whereas the scatter in (B) seems
to be more uniform.  Clearly, these data (with only eight points) are
not sufficient to draw any strong conclusions; nevertheless, they are
not inconsistent with either of the two possibilities $A_0\neq0$ or
$A_0=0$ in \aexp.

Therefore, it would be of considerable fundamental value to obtain a
confirmation (or otherwise) of the results of Praskovsky and Oncley,
with greater precision.  In particular, evidence that $A_0=0$ would
indicate that fully-developed turbulence is not a unique, Reynolds
number independent state, approached at large enough Reynolds number,
but instead is a state with universal Reynolds number and external scale
dependence, identical for different flows, but with no attainable
limit.

In conclusion, we have made two main points in this paper.  First, we
have suggested that there may not be complete similarity in Re --- in
particular, we have suggested that the lack of complete similarity is
exhibited as asymptotic covariance --- and we have detailed some of the
consequences of this.  In particular, a universal scaling should be
approached at large Re.  Second, the universal scaling may be that of
K41, indicating the existence of a unique fully-developed turbulent
state, or may have no asymptotic limit, indicating the non-existence of
such a simple state.

\ACK
We gratefully acknowledge helpful discussions with H. Aref, G. Eyink, A.
Majda, K. Sreenivasan and A. Yaglom.  We thank G. Eyink for permission
to quote from his unpublished work and A. Majda for bringing ref.
\prask\ to our attention after this work had been completed.  This work
was supported in part by National Science Foundation grant number
NSF-DMR-93-14938.

\refout
\figout

\end